\documentclass[12pt]{amsart}

\usepackage{natbib}
\usepackage{amscd,amssymb,amsmath,mathrsfs,amsfonts,graphicx,verbatim,epsfig,
psfrag, float, fancybox} 
\usepackage{graphics}                 
\usepackage{color}                    
\usepackage{hyperref}                 
\usepackage[TS1,OT1,T1]{fontenc}

\usepackage{amsmath}    
\usepackage{amscd,amssymb,mathrsfs,amsfonts}
\usepackage{graphicx}   
\usepackage{verbatim}   
\usepackage{color}      
\usepackage{amsmath}
\usepackage{verbatim}
\usepackage{setspace}
\usepackage{algorithm}

\begin{document}

\title{Comment on ``Detecting Novel Associations in Large Data Sets''
by Reshef et al, Science Dec 16, 2011}.

\author{Noah Simon} \author{Robert Tibshirani}

\thanks{Depts. of Statistics and Health Research and Policy, Stanford University.
Stanford, CA. 94305;  650-723-5989}

\maketitle

The proposal of \citet{reshef2011} is an interesting new approach for discovering non-linear dependencies among pairs of measurements in exploratory data mining. However, it has a potentially serious drawback. The authors laud the fact that MIC has no preference for some alternatives over others, but as the authors know, there is no free lunch in Statistics: tests which strive to have high power against all alternatives can have low power in many important situations. To investigate this, we ran simulations to compare the power of MIC to that of standard Pearson correlation and distance correlation (dcor). We simulated pairs of variables with different relationships (most of which were considered by the Reshef et. al.), but with varying levels of noise added. To determine proper cutoffs for testing the independence hypothesis, we simulated independent data with the appropriate marginals. As one can see from the Figure, MIC has lower power than dcor, in every case except the somewhat pathological high-frequency sine wave. MIC is sometimes less powerful than Pearson correlation as well, the linear case being particularly worrisome.

This set of dependencies is by no means exhaustive, however it suggests that MIC has serious power deficiencies, and hence when it is used for large-scale exploratory analysis it will produce too many false positives. The ``equitability'' property of MIC is not very useful, if it has low power. 

We believe that the recently proposed distance correlation measure of \citet{szekely2009} is a more powerful technique that is simple, easy to compute and should be considered for general use. A full R language script for our analysis appears in:\\
{\tt http://www-stat.stanford.edu/tibs/reshef/script.R}

\begin{figure}[t!]
  \centerline{
    \includegraphics[width=7in]{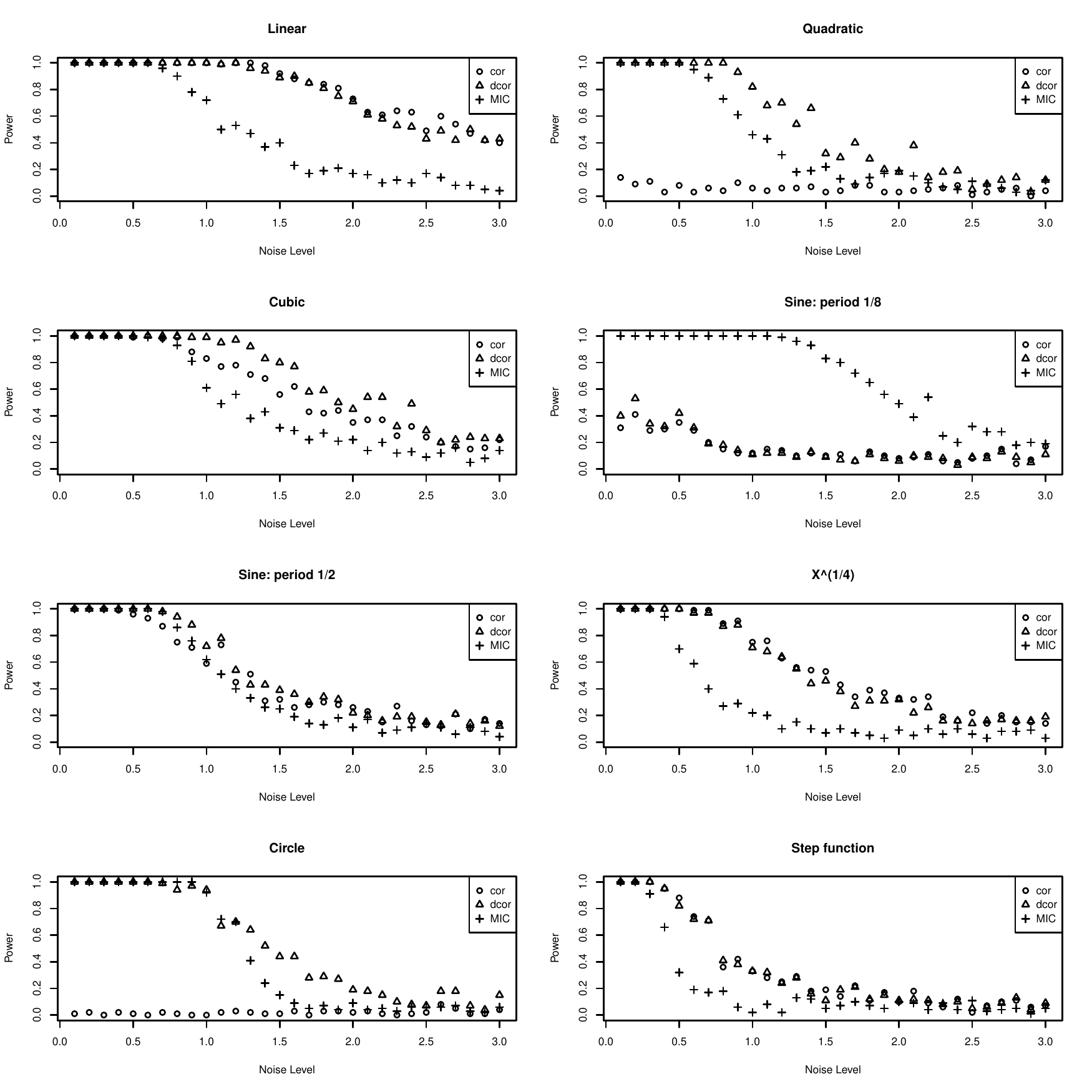}
  }
  \caption{Power of Pearson correlation (cor), distance
    correlation (dcor) and MIC  as a function of the level of noise added, in eight different scenarios.   The power is estimated via 500 simulations.
  } \label{fig:1} 
\end{figure}

\bibliographystyle{agsm}
\bibliography{texlib}

\end{document}